# Exploring the Efficacy of Convolutional Neural Networks in Sleep Apnea Detection from Single–Channel EEG

Chun Hin Siu
*AI and Computer Engineering*
*CMKL University*
Bangkok, Thailand
csiu@cmkl.ac.th

Hossein Miri
*International School of Engineering,*
*Faculty of Engineering*
*Chulalongkorn University*
Bangkok, Thailand
hossein.m@chula.ac.th

*Abstract* — Sleep apnea, a prevalent sleep disorder, involves repeated episodes of breathing interruptions during sleep, leading to various health complications, including cognitive impairments, high blood pressure, heart disease, stroke, and even death. One of the main challenges in diagnosing and treating sleep apnea is identifying individuals at risk. The current gold standard for diagnosis, Polysomnography (PSG), is costly, labor–intensive, and inconvenient, often resulting in poor–quality sleep data. This paper presents a novel approach to the detection of sleep apnea using a Convolutional Neural Network (CNN) trained on single–channel EEG data. The proposed CNN achieved an accuracy of 85.1% and a Matthews Correlation Coefficient (MCC) of 0.22, demonstrating a significant potential for home–based applications by addressing the limitations of PSG in automated sleep apnea detection. Key contributions of this work also include the development of a comprehensive pre–processing pipeline with an Infinite Impulse Response (IIR) Butterworth filter, a dataset construction method providing broader temporal context, and the application of SMOTETomek to address class imbalance. This research underscores the feasibility of transitioning from traditional laboratory–based diagnostics to more accessible, automated home–based solutions, improving patient outcomes and broadening the accessibility of sleep disorder diagnostics.

## I. INTRODUCTION

Sleep apnea is a sleep–disordered breathing condition characterized by repeated episodes of pauses in breathing or shallow breathing during sleep, leading to severe health complications, such as high blood pressure, heart disease, stroke, memory problems, depression, and anxiety (e.g., [1], [2], [3], [4]). Sleep apnea is classified into Central Sleep Apnea (CSA), caused by a failure of the brain to send signals to breathing muscles [5], and Obstructive Sleep Apnea (OSA), caused by physical airway blockage [6]. OSA is the most common type, affecting approximately 90% of all sleep apnea cases globally, translating into nearly 1 billion patients [2]. The gold standard for diagnosing sleep apnea is Polysomnography (PSG) which measures various physiological signals during sleep. PSG involves spending a night in a sleep laboratory, where multiple electrodes and sensors are attached to the patient's body to record brain activity, eye movements, muscle activity, and breathing. This setup is labor–intensive and costly, requiring skilled medical professionals to manually score the data. Due to socio–economic barriers and lack of awareness, PSG is often inaccessible to many individuals. Additionally, the *first night effect* where patients may have difficulty sleeping comfortably in a lab setting, can result in poor–quality sleep data and potential false–negative diagnoses [7]. This highlights the need for more accessible, cost–effective, and comfortable diagnostic methods.

Sleep apnea is identifiable through specific changes in EEG patterns that vary across sleep stages: wakefulness, REM, and the three NREM stages (N1, N2, N3) [8, 9]. During sleep apnea events, there is a noticeable decrease in beta band power (14–30 Hz) before the event, followed by an increase after it ends [10]. These EEG changes differ depending on the sleep stage. In NREM sleep, apneas show a gradual increase in delta–band activity (frequencies < 4 Hz), which decreases as the person awakens [10, 11]. In contrast, during REM sleep, apneas are marked by transient increases in delta–band activity and slight increases in beta band activity [10, 11]. REM sleep's tendency to promote muscle relaxation can also facilitate sleep apnea events [11]. Therefore, EEG, which reflects both sleep apnea events and sleep stages, is a valuable tool for detecting sleep apnea. The physiological connection suggests that Deep Learning with single–channel EEG, already successful in sleep stage classification, may also be effective for sleep apnea detection. Recent studies have demonstrated promising results in this area. For example, Hu et al. introduced a hybrid Transformer model with a self–attention mechanism, achieving a per–segment classification accuracy of 91% and an AUC of 0.96, outperforming other algorithms [12]. Mahmud et al. developed a hybrid CNN–Bi–LSTM (Convolutional Neural Network – Bidirectional Long Short Term Memory) approach using Variational Mode Decomposition (VMD) for EEG feature extraction, achieving over 89% average accuracy across multiple datasets with subject–independent cross–validation [13]. Additionally, Wang et al. designed a Bi–LSTM (Bidirectional Long Short Term Memory) model for sleep apnea detection, achieving 92.73% accuracy and 84.21% precision, optimized with Adam and 10–fold cross–validation [14]. These studies highlight the potential of Deep Learning models to enhance the accuracy and efficiency of sleep apnea detection using single–channel EEG, offering a convenient alternative to traditional methods. Challenges such as dataset imbalance, with normal breathing durations exceeding apnea events, complicate model evaluation. The MCC provides a balanced measure, considering true and false positives and negatives, ensuring accurate detection of both apnea and normal breathing events.

This study introduces a CNN–based method for automated sleep apnea detection using single–channel EEG data, with an accuracy of 85.1% and an MCC of 0.22. The comprehensive pre–processing and innovative dataset construction improve model performance by providing broader temporal context and addressing class imbalance. This research underscores the feasibility of transitioning from traditional laboratory–based diagnostics to more accessible, automated home–based solutions, improving patient outcomes and broadening the accessibility of sleep disorder diagnostics.

## II. METHODOLOGY

### A. Materials

The dataset used in this study is the St. Vincent's University Hospital / University College Dublin Sleep Apnea Database, available on PhysioNet [15]. This database comprises 25 full overnight polysomnograms with simultaneous three–channel Holter ECG recordings from adult subjects suspected of having sleep–disordered breathing. Subjects, aged between 28 and 68 years, with varying body mass indices as well as apnea–hypopnea indices, were selected from patients referred to the Sleep Disorders Clinic at St. Vincent's University Hospital. The polysomnograms recorded multiple signals, including EEG (C3–A2 and C4–A1, in 128 Hz), eye movements, muscle activity, and respiratory parameters. In this study, the C3–A2 channel was chosen according to the American Academy of Sleep Medicine recommendations [16]. Sleep stages and respiratory events were annotated by experienced sleep technologists in a 1–second resolution, providing a comprehensive dataset for developing and evaluating sleep apnea detection models.

### B. Pre–Processing EEG Signals with IIR Butterworth Filter

We utilized an Infinite Impulse Response (IIR) Butterworth filter to separate the EEG signal into five distinct frequency bands: delta (0.5–4 Hz), theta (4–8 Hz), alpha (8–12 Hz), sigma (12–16 Hz), and beta (16–40 Hz). Each band is associated with different brain activities and sleep stages, making it crucial to analyze them separately for accurate sleep apnea detection. The delta band is primarily linked to deep sleep, theta to light sleep and drowsiness, alpha to relaxed wakefulness, sigma to sleep spindles, and beta to active thinking and focus [17].

The Nyquist frequency, defined as half of the sampling frequency, represents the highest frequency that can be accurately sampled without introducing aliasing. In this dataset, the sampling frequency is 128 Hz, hence the Nyquist frequency is 64 Hz. Any frequency components above 64 Hz cannot be accurately captured or represented in the sampled signal. However, since the highest frequency band of interest is the beta band (16–40 Hz), this is well within the Nyquist limit, ensuring that all relevant EEG frequency bands can be accurately sampled and processed.

By filtering the signal into these bands, we can focus on the most relevant frequencies while excluding unwanted noise, such as the 60 Hz electrical noise commonly found in urban electric grids, as well as high–frequency noise from electronic devices, muscle artifacts, and low–frequency drift from electrode movement or sweating. After filtering, the signals from all bands are re–combined, ensuring that the EEG data retains its essential characteristics while being free from irrelevant noise. This pre–processing step enhances the clarity and reliability of the signal, enabling more accurate feature extraction and subsequent analysis by the CNN–based model.

### C. Z–Score Normalization

Following the filtering process, we applied Z–score normalization to the EEG data, transforming the data to have a mean of zero and a standard deviation of one. Z–score normalization is particularly important for Neural Network training as it ensures that features contribute equally to the learning process, enhancing the convergence speed and overall performance of the model. In the context of sleep apnea detection, Z–score normalization helps standardize the EEG signals across different subjects and sessions, making the model more robust and generalizable. This pre–processing step ensures that the CNN can effectively learn and distinguish patterns related to sleep apnea events from the normalized EEG signals. The comparison between raw data and pre–processed data is illustrated in **Figure 1**.

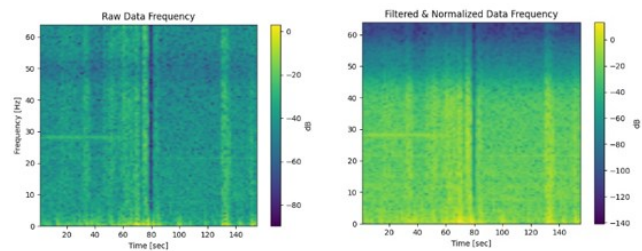

Fig. 1. Comparison of Raw Data and IIR Butterworth Filtered, Z–normalized Data

### D. Dataset Construction

For the construction of our dataset, the EEG data was segmented into 30–second intervals, as per the instructions in the International Classification of Sleep Disorders [20]. Each segment was assigned a label based on the professional annotations provided in the dataset. If a segment contained an apnea event – specifically labeled as 'APNEA–O' (obstructive), 'APNEA–C' (central), 'APNEA–M' (mixed), 'HYP–C' (hypopnea–central), 'HYP–O' (hypopnea–obstructive), or 'HYP–M' (hypopnea–mixed) – lasting longer than 10 seconds within the 30–second window, it was classified as positive for sleep apnea. Otherwise, it was classified as negative.

To ensure the model had sufficient contextual information, each 30–second segment was concatenated with 90 seconds of signals preceding and following it, resulting in a 210–second segment. This approach provided a broader temporal context to help the model accurately determine if an apnea event occurred within the current 30–second window [8]. However, the label assigned to each segment was based solely on the 30–second window itself, irrespective of the preceding and following signals. This method balances the need for detailed contextual information with precise labeling focused on the current segment.

### E. Data Splitting and Balancing

After constructing the dataset, we divided the data into training and validation sets using a 90–10 split, ensuring that the data was stratified by labels to maintain the same proportion of apnea and non–apnea segments in both sets. To address the inherent imbalance in the dataset, we applied SMOTETomek to the training data. SMOTETomek is a combination of Synthetic Minority Over–sampling Technique (SMOTE) and Tomek links. SMOTE increases the minority class representation (in this case, apnea segments) by generating synthetic samples, while Tomek links remove samples close to the decision boundary, often including noisy examples. Combining these techniques, SMOTETomek balances class distribution and cleans the data, improving training set quality. Applying SMOTETomek only to the training data ensures that the validation set reflects real–world distributions, providing accurate model performance assessment. This balancing step is crucial for enhancing the model's ability to detect under–represented apnea events effectively.

*F. Model Construction*

For the detection of sleep apnea, we designed a CNN model tailored to process the pre–processed EEG signals. The architecture of the CNN model is as follows (also shown in **Figure 2** – hyperparameters described in **Table 1**):

1. Input Layer: The model accepts input sequences of length corresponding to the 210–second segments with a single feature dimension representing the EEG signal.
2. Convolutional Layers: The model comprises four convolutional layers, each designed to extract different levels of features from the input data:
   - Convolutional Layer 1: This layer uses 8 filters with a kernel size of 35, followed by batch normalization, an Exponential Linear Unit (ELU) activation function, max–pooling with a pool size of 7, and a dropout rate of 0.1.
   - Convolutional Layer 2: This layer uses 128 filters with a kernel size of 175, followed by batch normalization, ELU activation, max–pooling with a pool size of 7, and a dropout rate of 0.1.
   - Convolutional Layer 3: This layer uses 16 filters with a kernel size of 175, followed by batch normalization, ELU activation, max–pooling with a pool size of 7, and a dropout rate of 0.1.
   - Convolutional Layer 4: This newly added layer uses 32 filters with a smaller kernel size of 3, followed by batch normalization, ELU activation, max–pooling with a pool size of 2, and a dropout rate of 0.1.
3. Fully Connected Layer: After flattening the output of the final convolutional layer, a dense layer with 64 units and ELU activation is applied, followed by a dropout layer with a rate of 0 to prevent overfitting.
4. Output Layer: The final layer is a dense layer with a single unit and a sigmoid activation function for binary classification, indicating the presence or absence of sleep apnea.

TABLE I. HYPERPARAMETERS USED IN THE PROPOSED CNN MODEL

| Layer | Hyperparameter | Value |
|---|---|---|
| Convolutional layer 1 | Kernel length | 35 |
|  | Number of filters | 8 |
| Convolutional layer 2 | Kernel length | 175 |
|  | Number of filters | 128 |
| Convolutional layer 3 | Kernel length | 175 |
|  | Number of filters | 16 |
| Convolutional layer 4 | Kernel length | 3 |
|  | Number of filters | 32 |
| MaxPooling layers | Window/stride size | 7,7,7,2 |
| Dense layer | Number of nodes | 64 |
| Convolutional layer dropout | Dropout rate | 0.1 |
| Dense layer dropout | Dropout rate | 0.0 |
| Optimizer | Learning rate | 0.0015 |

This CNN architecture is designed to be both efficient and effective, leveraging the convolutional layers to automatically extract meaningful features from the EEG signals, while the fully connected layer and output layer facilitate accurate classification of sleep apnea events. The use of batch normalization and dropout layers helps in regularizing the model, improving its generalization capabilities on unseen data. Overall, this tailored CNN model demonstrates a robust approach to detecting sleep apnea, ensuring reliable performance across various data samples. Through extensive testing, we have validated the model's ability to accurately identify apnea events, providing a promising tool for real–time, automated diagnosis.

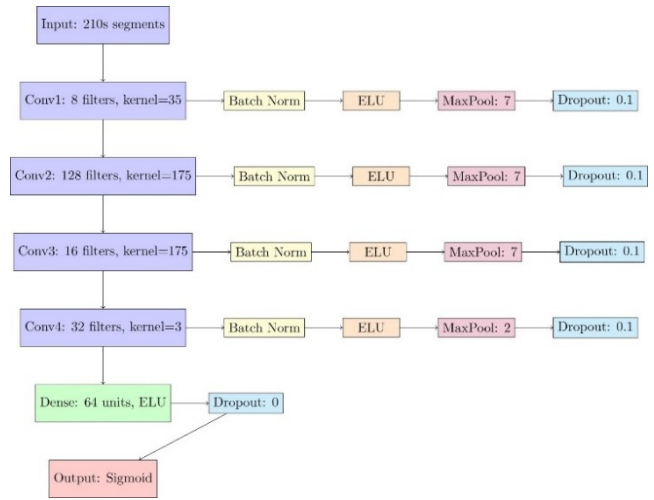

Fig. 2. Detail of the proposed CNN model

## III. RESULTS

The performance of the CNN model for sleep apnea detection using single–channel EEG data was evaluated using several metrics, including accuracy, confusion matrix, MCC, Receiver Operating Characteristic Area Under the Curve (ROC AUC), Cohen's Kappa, and F1 Score. These metrics provide a comprehensive overview of the model's ability to accurately detect sleep apnea events, considering the inherent imbalance in the dataset.

*A. Accuracy*

The validation accuracy achieved by our CNN model was 85.1%. This indicates that the model is capable of correctly classifying sleep apnea events in 85.1% of the validation cases. The training and validation accuracies are illustrated in **Figure 3** below.

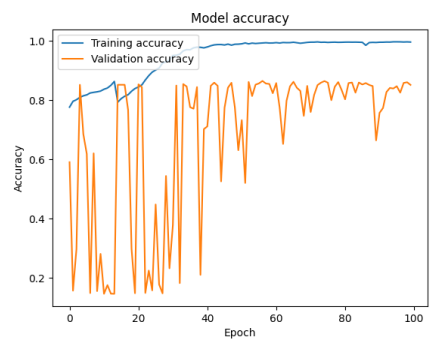

Fig. 3. Model Training and Validation accuracy

*B. Confusion Matrix*

The confusion matrix provides a detailed insight into the performance of the model, showing the distribution of true positives, true negatives, false positives, and false negatives. The confusion matrix is illustrated in **Figure 4** below. This matrix indicates that the model correctly identified 1682 instances of normal breathing (true negatives) and 88 instances of apnea (true positives). However, it also misclassified 94 instances of normal breathing as apnea (false positives) and 216 instances of apnea as normal breathing (false negatives).

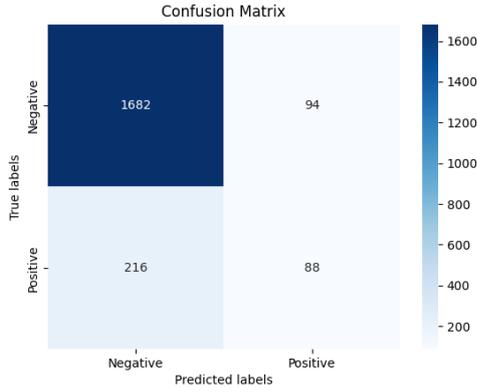

Fig. 4. Confusion Matrix of the Proposed Methodology

For comparison, the confusion matrix without using the oversampling technique is illustrated in **Figure 5**, with 1642 true negatives, 31 true positives, 134 false positives, and 273 false negatives.

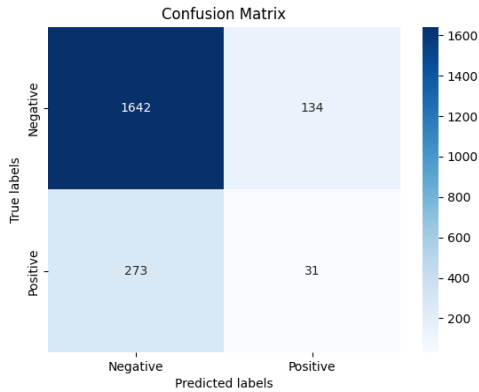

Fig. 5. Confusion Matrix of without using oversampling technique

### C. Matthews Correlation Coefficient (MCC)

The MCC for our model is 0.296, indicating a low–to–moderate correlation between the predicted and actual labels.

### D. ROC–AUC

The Receiver Operating Characteristic (ROC) curve's Area Under the Curve (AUC) for our CNN model was 0.618, indicating moderate performance. The training and validation ROC–AUC is illustrated in **Figure 6** below.

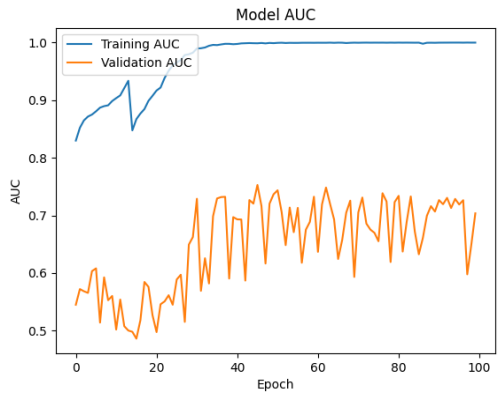

Fig. 6. Training and Validation ROC–AUC

## IV. DISCUSSION

Evaluating the performance of sleep apnea detection models presents significant challenges, primarily due to the natural imbalance in the dataset, where periods of normal breathing vastly outnumber apnea events. Traditional metrics, such as accuracy, can be misleading in such contexts. Our model achieved a validation accuracy of 85.1%, indicating its reliability for real–time, automated sleep apnea detection. However, the MCC of 0.296 and the ROC AUC of 0.618 indicate moderate performance. The confusion matrix revealed that while the model correctly identified many instances of normal breathing (1682 true negatives and 88 true positives), it also misclassified a significant number of events (94 false positives and 216 false negatives). This highlights the model's current limitation in detecting apnea events. The application of the SMOTETomek oversampling technique improved the true positive rate from 31 to 88 and reduced false negatives from 273 to 216, indicating enhanced sensitivity and reliability. Cohen's Kappa value of 0.2837 and the F1 Score of 0.3621 reflect the model's moderate agreement and balance between precision and recall, respectively. These results underscore the need for further refinement to improve the model's sensitivity to apnea events and its overall generalizability. To address the limitations observed in this study, several key areas for future work will be considered:

- *Enhanced Neural Network Architectures*: Exploring hybrid Neural Network architectures, such as combining CNNs with RNNs or Transformers, could improve apnea detection by capturing spatial and temporal dependencies in EEG signals. However, performance enhancements must be balanced with the computational limitations of mobile devices to ensure models remain efficient and practical for real–time use.

- *Incorporation of Additional Physiological Signals*: Incorporating additional physiological signals, such as heart rate variability or oxygen saturation, with EEG data could enhance the model's accuracy and performance, particularly in distinguishing between different types of sleep apnea.

- *Extensive Validation on Diverse Datasets*: Extensive validation on larger, diverse datasets is crucial for ensuring the model's applicability across various populations and settings. Collaborations with clinical institutions for real–world data and feedback from medical professionals will also help refine the model and its practical deployment.

## V. CONCLUSION

Our proposed CNN-based approach for sleep apnea detection using single-channel EEG data demonstrates promising results, providing a valuable preliminary tool for initial screening. While this method offers insights and holds potential, it is essential to acknowledge its limitations and recognize the need for further research and development to achieve the ultimate goal of accessible, efficient, and reliable sleep apnea diagnostics. For definitive diagnoses in critical cases, comprehensive PSG in a sleep lab setting remains indispensable. Through innovation and collaboration, we can strive to overcome these limitations and harness the full potential of this technology to improve patient outcomes and broaden the accessibility of sleep disorder diagnostics.